\documentclass[12pt]{article}
\newlength{\abstractwidth}
\usepackage{epsf}
\usepackage{color}
\usepackage{graphicx}
\usepackage{amssymb}

\renewcommand{\thefootnote}{\fnsymbol{footnote}}
\renewcommand{\thanks}[1]{\footnote{#1}}
\newcommand{\starttext}{
\setcounter{footnote}{0}
\renewcommand{\thefootnote}{\arabic{footnote}}}

\newcommand{\bea}{\begin{eqnarray}}
\newcommand{\eea}{\end{eqnarray}}
\newcommand{\ee}{\end{equation}}
\newcommand{\be}{\begin{equation}}

\newcommand{\sm}{\smallskip}

\def\cA{{\cal A}}

\def\cF{{\cal F}}

\def\cN{{\cal N}}

\def\tr{{\rm tr}}

\def\half{ {1\over 2}}
\def\p{\partial}

\def\l({\left(}
\def\r){\right)}

\def\a{\alpha}
\def\b{\beta}

\def\g{\gamma}

\def\g{\gamma}

\def\Bc{{\cal B}}

\def\xt{\tilde{x}}
\def\yt{\tilde{y}}

\def\no{\nonumber}

\begin{document}
\starttext
\setcounter{footnote}{0}

\bigskip

\begin{flushright}
\today
\end{flushright}

\bigskip

\vskip 2cm

\begin{center}

{\Large \bf Magnetic Brane Solutions in AdS }

\medskip

\vskip .4in

{\large \bf Eric D'Hoker$^{1}$ and  Per Kraus$^{1,2}$}

\vskip .2in

\centerline{\it ${}^{1}$ Department of Physics and Astronomy, UCLA,
Los Angeles, CA 90095-1547, USA}

\vskip.2in

\centerline{\it ${}^{2}$ Kavli Institute for Theoretical Physics, UCSB, Santa Barbara, CA 93106, USA}

\vskip.2in

{\tt \small dhoker@physics.ucla.edu; pkraus@ucla.edu}


\end{center}

\vskip .2in

\begin{abstract}

\vskip 0.1in

We construct asymptotically  AdS$_5$ solutions of Einstein-Maxwell theory dual to $\cN=4$ SYM theory on $\mathbb{R}^{3,1}$ in the presence of a background magnetic field. The solutions
interpolate between AdS$_5$ and a near horizon AdS$_3\times T^2$.  The central charge of the near horizon region, and hence low temperature entropy of the solution,
is found to be $\sqrt{4\over 3}$ times that of free $\cN=4$ SYM theory.  The entropy vanishes
at zero temperature.   We also present
the generalization of these solutions to arbitrary spacetime dimensionality.

\end{abstract}

\newpage


\newpage

\section{Introduction}
\setcounter{equation}{0}

In the AdS/CFT correspondence  boundary gauge theories in the presence of background electromagnetic fields can be studied by imposing suitable boundary conditions on gauge fields in the bulk of AdS.  By turning on such fields one can compute the electrical
conductivity of the gauge theory, study its response to magnetic fields, and so forth.

The solution for a black brane in AdS$_4$ with a magnetic field is easily found and has
many AdS/CFT applications   (e.g \cite{Hartnoll:2007ai}-\cite{Basu:2009qz}) to the study
of $2+1$ gauge theories in magnetic fields.  It is clearly  of interest to have such solutions in the AdS$_5$ case as well.   For instance, as recently emphasized in \cite{Son:2009tf},
strongly coupled gauge theories in magnetic fields arise at RHIC, and one would like to
be able to use holography to study the effect of the magnetic field.  To the extent
that $3+1$ dimensional condensed matter systems can be modeled by AdS/CFT, the ability to
turn on magnetic fields  provides a valuable probe of the system with a clear
physical meaning.

Given this motivation, in this paper we find magnetic brane solutions to five dimensional  Einstein-Maxwell theory with a negative cosmological constant.   The most important
properties of these solutions can be determined analytically, although some numerical  work
is needed to capture all the details.

The solutions that we find  are related to, but distinct from, previously studied  ``AdS$_5$ black string" solutions (and their higher dimensional cousins)   \cite{Chamseddine:1999xk}-\cite{Bernamonti:2007bu}, as well as the solutions of Maldecena and Nunez \cite{Maldacena:2000mw}.  In these papers, two of the spatial direction are taken
to be compact (usually $S^2$ or $H^2$) and the magnetic field strength is taken to be proportional to the curvature two-form, as is in fact required by supersymmetry.  On the other hand, we will be looking for solutions
with a nonzero field strength on a flat boundary metric.    Our solutions are thus intrinsically non-supersymmetric.

Taking the spatial boundary directions to be a compact torus, our solutions interpolate
between  AdS$_3 \times T^2$ at small $r$ and AdS$_5$ at large $r$.   At finite temperature
the AdS$_3$ factor is replaced by a BTZ black hole \cite{Banados:1992wn}.   The black
brane entropy density correspondingly  interpolates between a linear $T$ dependence at low temperature and a $T^3$ dependence at high temperature.

We then compare our results to the thermodynamics of free $\cN=4$ SYM theory in an external
 $U(1)_R$ magnetic field.   At high temperature we recover the standard result
$S_{\rm grav} = {3\over 4}S_{\cN=4}$.   At low temperature the $\cN=4$ theory reduces to a
$1+1$ dimensional conformal field theory described by the fermion zero modes, with a central
charge equal to $N^2$ times the number of units of quantized magnetic flux.  The corresponding strong coupling result can be computed in gravity from the Brown-Henneaux formula, and we
find   an increase by a factor of $\sqrt{4\over3}$,  i.e. $c_{\rm grav} =\sqrt{4\over 3}~c_{\cN=4}$.  The low temperature entropy computed from gravity therefore is also enhanced
by this factor compared to the free $\cN=4$ result.   It is amusing that turning on a
magnetic field actually improves the numerical agreement between gravity and free $\cN=4$ SYM, giving a relative $\sqrt{4\over 3}$ versus  ${3\over 4}$.

The property that the entropy of the AdS$_5$ magnetic brane solution vanishes at zero temperature distinguishes it from the AdS$_4$ case, where there is a finite entropy at
extremality.   In fact, this property also matches what one would expect from a free
fermion description.  In the AdS$_4$ case all spatial directions of the boundary are
threaded by magnetic flux; there is thus a finite density of fermion zero modes
per unit area, and hence a finite entropy density.

It is straightforward to extend our considerations to arbitrary spacetime dimension.
For odd dimensional AdS$_{d+1}$ spacetimes and maximal rank magnetic field, we find
solutions interpolating between AdS$_3 \times T^{d-2}$ and AdS$_{d+1}$, with properties
very similar to the AdS$_5$ case.  Similarly, for even dimensional AdS$_{d+1}$ spacetimes
the situation parallels the AdS$_4$ case; explicit solutions are easily found, and the
solutions have a finite entropy extremal limit.

This paper is organized as follows. In section 2 we construct magnetic brane solutions
in AdS$_5$.  In section 3 we compute the thermodynamics of free $\cN=4$ SYM theory
in a background magnetic field.  In section 4 we generalize to arbitrary spacetime
dimension. Some conclusions are given in section 5.   The appendix details our progress
in looking for analytic solutions of the AdS$_5$ equations.

\section{Magnetic brane in AdS$_5$}
\setcounter{equation}{0}

The action of five-dimensional Einstein-Maxwell theory with a negative cosmological constant is\footnote{Conventions:  $R^\lambda_{~\mu\nu\kappa}= \p_\kappa \Gamma^\lambda_{\mu\nu}-\p_\nu \Gamma^\lambda_{\mu\kappa} +\Gamma^\eta_{\mu\nu}\Gamma^\lambda_{\kappa\eta}-\Gamma^\eta_{\mu\kappa}\Gamma^\lambda_{\nu\eta}$
and $R_{\mu\nu} = R^\lambda_{~\mu \lambda \nu}$.  }
\bea\label{ba}
S&=& -{1\over 16\pi G_5}\int\! d^5 x\sqrt{-g}\Big( R + F^{MN}F_{MN}-{12\over L^2} \Big)+S_{\rm{bndy}}~.
\eea
The boundary terms include the Gibbons-Hawking term as well as other contributions necessary
for a well posed variational principle \cite{Henningson:1998gx,Balasubramanian:1999re}; their explicit forms will not be needed here.
Along with the Bianchi identity, the field equations are
\bea\label{bb}
R_{MN} &=& { 4\over L^2} g_{MN} +{1 \over 3} F^{PQ}F_{PQ} g_{MN} -2 F_{MP}F_N^{~P} \\ \no
\nabla^M F_{MN} & = & 0~.
\eea
 We henceforth set the AdS radius to unity: $L=1$.

Were we to add to the action the Chern-Simons term
\bea\label{bc}
S_{CS} = {k\over 16\pi G_5} \int  A\wedge F\wedge F~,\quad \quad k ={8\over 3\sqrt{3}}
\eea
then our action would correspond to  the bosonic part of $D=5$ minimal  gauged supergravity (e.g. \cite{Gauntlett:2003fk}).  The Chern-Simons term makes no contribution to the solutions
considered in this paper, but will be useful for fixing the normalization of the gauge field.

We are interested in solutions asymptotic to AdS$_5$  with a magnetic field tangent to the boundary directions.  The field strength and metric can be taken to be invariant under
spacetime translations, rotations in the $x^{1,2}$ plane, and time reversal.   A general ansatz consistent with the symmetries is
\bea\label{bd}
ds^2 & = & -U(r)dt^2 + {dr^2 \over U(r)} + e^{2V(r)}\Big( (dx^1)^2+ (dx^2)^2 \Big) + e^{2W(r)}dy^2 \\ \no
F & = & B dx^1 \wedge dx^2
\eea
The Maxwell equation is automatically satisfied, and the Einstein equations reduce to
\bea
\label{bda}
U(V''-W'') + \Big ( U' +  U(2 V'+W') \Big ) (V'-W') & = & - 2 B^2 e^{-4V}
\\
2 V'' + W'' + 2(V')^2  +(W')^2 & = & 0
\no \\
\half U'' + \half U'(2 V' + W') & = &
	 4  + {2  \over 3} B^2 e^{-4V}
\no \\
2U'V'+U'W' + 2 U (V')^2 + 4 U V'W'
& = &  12   - 2   e^{-4V}  B^2
\no
\eea
With $r$ as the evolution parameter, the final equation represents a constraint on initial data.  Once imposed on the initial data, it is automatically satisfied on the full solution by virtue of the three dynamical equations.   Alternatively, one can omit one of the dynamical
equations, since it will be implied by the remaining dynamical equations together with
the derivative of the constraint equation.

With $B=0$, a solution to these equations is AdS$_5$, represented by $U=e^{2V}=e^{2W}=r^2$.
For nonzero $B$ an exact solution is given by $U=3(r^2-r_+^2)$, $e^{2V}= B/ \sqrt{3}$,
$e^{2W}=3r^2$.    This solution represents the product of a BTZ black hole and $T^2$ (we are taking   the $x^{1,2}$ directions to be compact),
\bea\label{be}
ds^2= -3(r^2-r_+^2)dt^2+{dr^2 \over 3(r^2-r_+^2)}+ {B\over \sqrt{3}}\Big( (dx^1)^2+ (dx^2)^2 \Big) +3r^2 dy^2~.
\eea
Our goal is to find solutions
that interpolate between (\ref{be}) at small $r$ and AdS$_5$ at large $r$.   From the boundary field theory point of view this represents an RG flow between a $D=3+1$ CFT at short distance and a $D=1+1$ CFT at long distance.   As we discuss  in the next section, this is the expected behavior of $\cN=4$ SYM in the presence of uniform external magnetic flux.

The central charge of the near horizon AdS$_3$ region can be computed from the Brown-Henneaux
formula \cite{Brown:1986nw}, $c=3l/ (2G_3)$.   The AdS$_3$ radius is $l = 1/\sqrt{3}$,   Taking $x^{1,2}$ to be compact with coordinate volume $V_2$, the $D=3$ Newton constant is $G_3 = \sqrt{3}G_5/ (B V_2)$.
Using the AdS$_5$/CFT$_4$ relation $G_5 = \pi /( 2N^2)$, we find
\bea\label{bf}
c  = {N^2 B V_2 \over \pi} = \sqrt{4\over 3} \left({\Bc V_2 \over 2\pi}\right)N^2~.
\eea
In the last step we have written the result in a form convenient for comparison with
$\cN=4$ SYM theory, where the rescaled magnetic field is $\Bc= \sqrt{3}B$.  The combination
$\Bc V_2/( 2\pi)$ will be identified with the number of quantized units of
magnetic flux.    Knowledge of the central charge (\ref{bf}) will be sufficient information
to deduce the low temperature behavior of the black hole entropy.

We have not succeeded  in solving (\ref{bda}) analytically to find the
interpolating solutions, but numerical integration is straightforward.   The procedure is a bit  different depending on whether the temperature is zero or nonzero.

\subsection{Zero temperature solutions}

At zero temperature we can look for solutions preserving Lorentz invariance in the
$(t,y)$ directions, which corresponds to setting $U=e^{2W}$.  The system of equations (\ref{bda}) reduces to
\bea\label{bg}
&& 2 V'' + W'' + 2(V')^2  +(W')^2  =  0 \\ \no
&& V'^2 +W'^2 +4V' W' = 6e^{-2W}-e^{-4V-2W}B^2~.
\eea
Note that $B$ can be absorbed by a shift of $V$.
Starting from the small $r$ behavior
$e^{2V} = B/\sqrt{3}$ and $e^{2W} = 3r^2$,
we numerically integrate out and find a
solution with large $r$ behavior $e^{2V}= v r^2$ and $e^{2W}=r^2$.
   The result is a smooth zero temperature solution interpolating between
a near horizon AdS$_3 \times T^2$ and an asymptotic AdS$_5$.  There is a unique such
solution, inasmuch as the value of $B$ can be absorbed by a  rescaling  of $x^{1,2}$.

\subsection{Finite temperature solutions}

Now we turn to the finite temperature solutions.  The solutions carry a nonzero temperature
and magnetic field; however, using the freedom to rescale coordinates there is really only
a one parameter family of solutions, which we can think of as being parameterized by
the dimensionless combination ${T\over \sqrt{B}}$.    The numerical analysis proceeds
as follows.  We rescale $r$ such that the horizon is at $r=1$; i.e. $U(1)=0$.
By rescaling $t$ we can take $U'(1)=1$, which sets the temperature to a fixed value, leaving $B$ as the free parameter.   Further, by rescaling $x^{1,2}$ and $y$ we can
take $V(1)=W(1)=0$.    With these conditions, the first and third equations of (\ref{bda})
give us the initial data $V'(1)=4-{4\over 3}b^2$ and $W'(1)=4+{2\over 3}b^2$, where we are writing $b$ for the value of the magnetic field in these coordinates.      We then
integrate out to find solutions with asymptotic behavior $U=r^2$, $e^{2V}=vr^2$, and
$e^{2W}=wr^2$,  where $v$ and $w$ are functions of the free parameter $b$, to be computed
numerically.   We find smooth solutions for  $b<\sqrt{3}$, thus exhibiting the
existence of solutions interpolating between a  near horizon BTZ $\times T^2$ and an
asymptotic AdS$_5$.

The solution has a conformal boundary metric $ds^2 = -dt^2 + v \Big((dx^1)^2+(dx^2)^2\Big) + w dy^2$.   To put this in standard form we can introduce the coordinates $\xt^{1,2} =\sqrt{v}x^{1,2}$ and $\yt = \sqrt{w}y$.  The Hawking temperature, determined from the imaginary time periodicity, is $T = 1/( 4\pi)$.
Since the field strength takes the form
\bea\label{bga}
F= bdx^1 \wedge dx^2 = {b\over v}d\xt^1 \wedge d\xt^2~,
\eea
it is $B=b/ v$ that represents the  physical magnetic field.   Similarly,
the  physical entropy density is
\bea\label{bgb}
{S\over V} = {1\over 4G_5}{1\over v\sqrt{w}}~.
\eea

It is most illuminating to display the numerical results as a plot of the entropy density versus temperature.
Since it is only dimensionless quantities that are meaningful, we divide each by the
appropriate power of the magnetic field $B$.   Or rather since it is $\Bc =\sqrt{3}B$ that will appear naturally on the field theory side, we divide by powers of $\Bc$.
For the temperature, we thus compute
\bea\label{bgc}
{T\over \sqrt{\Bc}} = {3^{-1/4}\over 4\pi}\sqrt{v\over b}~.
\eea
For the entropy density, using $G_5 = {\pi \over 2N^2}$ and dividing through by
$N^2$ we compute
\bea\label{bgd}
{S\over V N^2 \Bc^{3/2}} = {3^{-3/4}\over 2\pi} \sqrt{v\over b^3 w}~.
\eea

\begin{figure}[h!!t]
\begin{centering}
\includegraphics[scale=0.6]{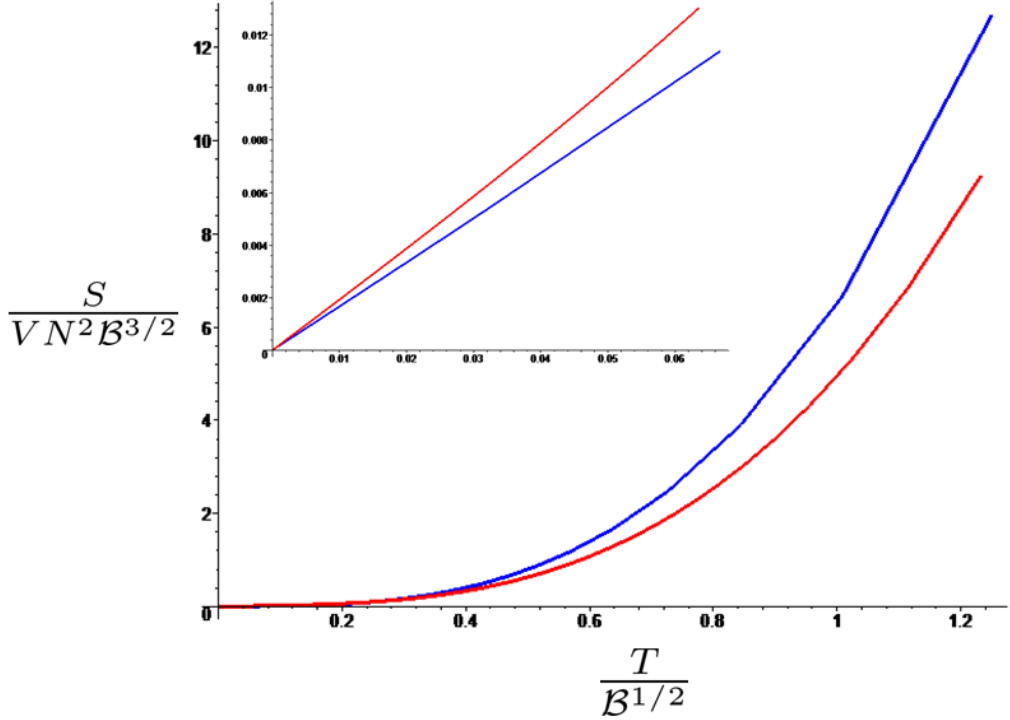}
\caption{Plot of entropy versus temperature for gravity (red) and free $\cN=4$ SYM theory (blue). The  inset shows the low temperature behavior.   At low temperature the entropies are linear in $T$, with $S_{\rm grav} = \sqrt{4\over 3}S_{\cN=4}$.   At high temperature the entropies are cubic in $T$, with $S_{\rm grav} = {3\over 4}S_{\cN=4}$.  Gravity gives
the larger entropy at low temperature by a factor of $\sqrt{4\over 3}$; as the temperature
is raised the curves cross, and then asymptotically the gravitational entropy is lower
by a factor of ${3\over 4}$.   }
\label{Scompare}
\end{centering}
\end{figure}

The numerical results, together with those of the field theory computation discussed in
the next section, are shown in Figure \ref{Scompare}.

The behavior of the black brane entropy at high and low temperatures can be determined analytically.  At high temperatures the magnetic field becomes a subleading effect,
and we recover the standard result for finite temperature D3-branes, namely a $T^3$
dependence with the entropy being $3/4$ of that obtained from  the free field limit of
$\cN=4$ SYM theory with gauge group $U(N)$.

At low temperature we have a BTZ black hole.  The entropy of a BTZ black hole (or more generally any system with the symmetries of a $D=1+1$ CFT) is
given by $S = {\pi \over 3}cTL_y$, where we are taking the $y$ coordinate to be compact, and
where in general $c$ is the average of the left and right moving central charges.     One might be worried that since $g_{yy}$ is a nontrivial function of $r$,  the relevant size of the $y$ circle differs depending on whether
we measure it in the BTZ region or at infinity in the AdS$_5$ region.  The same
is true of the time coordinate, which determines the temperature.  Fortunately, due
to the zero temperature  condition $U=e^{2W}$, the two possible rescalings cancel, so that
the product $TL_y$ is the same in the BTZ region as at infinity.   Using the result (\ref{bf}) the low temperature entropy becomes
\bea\label{bh}
{S\over V} = {N^2 \over 3\sqrt{3}}\Bc T~.
\eea
We have verified that (\ref{bh}) indeed matches the low temperature numerics, as can be
seen from the inset in  Figure \ref{Scompare}.

The numerics show that the entropy smoothly interpolates between the linear and cubic in
$T$ dependence as the solution interpolates between a BTZ black hole and a five dimensional
black brane.

\section{Comparison with ${\cal N}=4$ Super Yang-Mills}
\setcounter{equation}{0}

We now work out the entropy of the free field limit of $\cN=4$ SYM theory in an external magnetic field in order to compare with the black brane entropy.    Our first task is to fix
the normalization of the field theory magnetic field relative to that used on the gravity side.  We should first state more specifically which magnetic field is under consideration.
On the gravity side, we stated that if we include the term (\ref{bc}) then the action
corresponds to minimal gauged supergravity.  This implies that the bulk gauge field appears
on the boundary as an external field coupled to the R-symmetry current of the field theory.
Here we are thinking of the $\cN=4$ theory  in $\cN=1$ terms, with a $U(1)$ R-symmetry.
The natural normalization in the field theory is to assign the gaugino R-charge $1$.
In this normalization, the field content of the $\cN=4$ theory is as follows:  we have
$N^2$ Weyl spinors of charge $1$ (the gauginos);  $3N^2$ Weyl spinors of charge $-{1\over 3}$ (the fermions in the chiral multiplets);  $3N^2$ complex scalar fields of charge ${2\over 3}$ (the scalars in the chiral multiplets); and $N^2$ vector fields of charge $0$ (the gauge fields).

To fix the relative normalizations we compare the anomalous variations under gauge transformations of the R-symmetry gauge fields, $\delta \cA = d\tilde{\Lambda}$.   On the field theory side we use
the standard result from the triangle anomaly,
\bea\label{bha}
\delta S_{eff} = {1\over 24\pi^2}\tr Q^3 \int\! \tilde{\Lambda} \cF\wedge \cF~,
\eea
where the trace is over the spectrum of Weyl fermions.   In our case, $\tr Q^3 = {8N^2 \over 9}$, and so
\bea\label{bhb}
\delta S_{eff} ={N^2\over 27\pi^2} \int\! \tilde{\Lambda} \cF\wedge \cF~.
\eea
On the gravity side the anomalous variation comes from the Chern-Simons term (\ref{bc}), whose coefficient was fixed by supersymmetry.   This gives
\bea\label{bhc}
\delta S = {k \over 16\pi G_5}\int\!
\Lambda F \wedge F = {N^2 \over 3^{3/2} \pi^2}\int\! \Lambda F \wedge F~.
\eea
Comparing, we find the relation $\cF = \sqrt{3} F$, which is the result that we used in the previous section.

We now compute the partition function, $Z={\rm Tr}e^{-\beta H}$, of $\cN=4$ SYM theory at finite temperature and magnetic field,  at zero coupling.   From the partition function
we extract the entropy using the standard formula $S=(1-\beta {\p \over \p \beta})\ln Z$,
and then compare with the black brane entropy.
For $\Bc=0$ it is well known that the two entropies only differ by a factor of ${3\over 4}$,
even though the field theory and gravity computations are valid in the non-overlapping
regimes of small and large 't Hooft coupling.  It is interesting to extend this comparison
to nonzero magnetic field; as we'll see, this actually improves the agreement between the entropies.

In the free field limit, to compute $Z$ we only need to know the spectrum  of single particle excitations in the presence of a magnetic field pointing along the $y$ direction, which are given by relativistic Landau levels.       First consider a charge $q_\phi$ scalar field.  Solving $D^\mu D_\mu \phi =0$, we find the energies
\bea\label{bi}
E=\pm \sqrt{p_y^2 +(2n+1)|q_\phi \Bc|}~,\quad\quad n=0,1,2,\ldots
\eea
As usual, the branch with $E<0$ corresponds to charge $-q_\phi$ antiparticles with positive energy.   Each mode has a degeneracy corresponding to the number of units of magnetic flux,
$|q_\phi \Bc|V_2 /( 2\pi)$, where $V_2$ denotes the area in the $x^{1,2}$ plane
transverse to $\vec{\Bc}$.   Summing over both the particles and anti-particles, the scalar
partition function is
\bea\label{bj}
\ln Z_\phi(q_\phi) &=&-2{ |q_\phi\Bc| V_2 \over 2\pi} \sum_{n=0}^\infty {L_y \over 2\pi}\int_{-\infty}^\infty dp_y   ~\ln\Big(1-e^{-\beta\sqrt{p_y^2+ |q_\phi\Bc|(2n+1)}}~\Big)~.
\eea

Next, consider a charge $q_\psi$  Weyl spinor.  We solve $\gamma^\mu D_\mu \psi=0$ subject
to $\gamma^5 \psi = \psi$.  This yields the following spectrum of energies.  First, there
are solutions obeying $E^2 > p_y^2$ with
\bea\label{bk}
E = \pm \sqrt{p_y^2 +2|q_\psi \Bc| n}~,\quad\quad n =1,2,\ldots
\eea
Second, there are solutions with $E^2 = p_y^2$.    For $q_\psi \Bc >0$ these obey $E=p_y$;
for $q_\psi \Bc <0$ they obey $E=-p_y$.    These are zero modes of the two-dimensional Dirac
operator, whose existence is mandated by the index theorem.  The sign of the momentum of
the physical  $n=0$ excitations is correlated with the sign of $q_\psi$.
All of the solutions have degeneracy
$|q_\psi \Bc| V_2/ (2\pi)$.    The partition function of a charge $q_\psi$ Weyl spinor
is thus\footnote{For a given sign of $q_\psi$ the zero mode fermions really have a definite
sign of $p_y$, but to simplify (\ref{bl}) we replace the integration over anti-particles
of $p_y>0$ (say) by the equivalent integration over $p_y<0$.}
\bea\label{bl}
\ln Z_\psi(q_\psi) &=&{ |q_\psi\Bc| V_2 \over 2\pi}  \sum_{n=0}^\infty\sum_{\alpha=\pm 1}{L_y \over 2\pi}\int_{-\infty}^\infty  dp_y   ~\ln\Big(1+e^{-\beta\sqrt{p_y^2+ |q_\psi\Bc|(2n+1-\alpha)}}~\Big)~.
\eea
The $\alpha=1$ part includes the zero mode contribution.

The gauge fields are neutral, and have partition function
\bea\label{bm}
\ln Z_V = -2{V_2 L_y \over (2\pi)^3}\int \! d^3p \ln\Big(1-e^{-\beta|\vec{p}|}\Big)~.
\eea

The total partition function corresponding to the field content of $\cN=4$ SYM is
\bea\label{bn}
\ln Z =N^2\Big( 3\ln Z_\phi(2/3) + \ln Z_\psi(1)+3\ln Z_\psi(-1/3) + \ln Z_V\Big)~.
\eea

In the high temperature limit the sums over $n$ can be replaced by integrals, and we
recover the standard result for the entropy,
\bea
{S\over V} ={2\pi^2 \over 45}\Big(g_b+{7\over 8}g_f\Big)T^3
\eea
where $g_{b,f}$ denote the number  of bosonic/fermionic helicity states.
In the present context, $g_b = g_f = 8N^2$.

In the low temperature limit the partition function is dominated by the fermionic zero
mode contribution.   From (\ref{bl}) we see that each $D=3+1$  fermion contributes
the same as $|q_\psi \Bc| V_2/ (2\pi)$  fermions in $D=1+1$, with corresponding
central charge $c= |q_\psi \Bc| V_2/ ( 4\pi)$.  The low temperature
entropy is thus
\bea\label{bo}
S\approx  {\pi \over 3}c L_y T~,\quad\quad c= \sum_\psi  {1\over 2} {|q_\psi \Bc| V_2 \over 2\pi}
=  {| \Bc| V_2 \over 2\pi} N^2~.
\eea
Comparing with (\ref{bf}) we see that the central charges, and hence the low temperature
entropies, differ by a factor of $\sqrt{4\over 3}$.   Somewhat surprisingly, the strong
coupling result coming from gravity gives the  larger central charge, in contrast
to the result that at high temperature the gravitational entropy is less by a factor of
$3/4$ compared to free $\cN=4$ SYM theory.

For intermediate values of the temperature we evaluate the sums and integrals numerically, and obtain the result displayed in Figure \ref{Scompare}.  Both the gravity and field
theory entropies smoothly interpolate between a linear and cubic temperature dependence,
corresponding to the fact that both are interpolating between $D=1+1$ and $D=3+1$ CFTs.

\section{Generalization to AdS$_{d+1}$}
\setcounter{equation}{0}

In this section we discuss black brane solutions with magnetic fields in arbitrary
dimensions.   There are two basic cases, depending on whether the spacetime dimensionality is odd or even.   The odd dimensional case, with maximal rank B-field, is analogous to the
solutions constructed in section 2. The solutions interpolate between a near horizon AdS$_3$ and an asymptotic AdS$_{d+1}$.  The solutions in the even dimensional case are analogous
to the magnetic black brane in AdS$_4$ (see \cite{Ortaggio:2007hs} for related solutions).  These black branes have an extremal limit with
nonzero entropy density, and a corresponding near horizon AdS$_2$ factor.  For both
odd and even dimensions, the low temperature behavior of the entropy matches that of massless free fermions in magnetic fields.

More generally, by considering B-fields of less than maximal rank, one can have solutions
interpolating between a near horizon AdS$_{d+1-2r}$ and an asymptotic AdS$_{d+1}$, where
$r$ is a positive integer.

In $d+1$ dimensions the Einstein-Maxwell equations are
\bea
\label{ca}
R_{MN} &=& { d\over L^2} g_{MN} +{1 \over d-1} F^{PQ}F_{PQ} g_{MN} -2 F_{MP}F_N^{~P} \\ \no
\nabla^M F_{MN} & =& 0
\eea
There is also the Bianchi identity. For clarity, we restore the $L$-dependence in this section.

\subsection{$d$ odd}

We choose a field strength of maximal rank. By rotating coordinates we can skew-diagonalize
$F_{MN}$, and for simplicity we restrict to the case of equal skew-eigenvalues,
 $F_{12}=F_{34}=\ldots =B$.   The metric ansatz is
\bea\label{cb}
ds^2 = -U(r)dt^2 +{dr^2\over U(r)} + e^{2V(r)}\Big((dx^1)^2 + \cdots + (dx^{d-1})^2\Big)~.
\eea
The Einstein equations become
\bea\label{cc}
&& U'' +(d-1)U'V'={2d\over L^2}+2B^2 e^{-4V} \\ \no
&&    U''+2(d-1)\Bigg(UV''+UV'^2+{U'V'\over 2}\Bigg)  ={2d\over L^2}+2B^2 e^{-4V} \\ \no
&&U V''+ U'V'  +(d-1)UV'^2 = {d\over L^2}-B^2 e^{-4V}
\eea
A solution is given by
\bea\label{cd}
U = {r^2 \over L^2} +\left({L^4 \over 4-d}\right)  {B^2 \over r^2}  -{M\over r^{d-2} }~,\quad\quad e^{2V}={r^2 \over L^2}~.
\eea
This is a magnetic black brane, with a horizon at $U(r_+)=0$.

The Hawking temperature is
\bea\label{ce}
T= {1\over 4\pi} U'(r_+) = {1\over 4\pi}\Bigg( {dr_+ \over L^2} - {L^4 B^2\over  r_+^3} \Bigg)~.
\eea
The extremal limit is thus  $r_+^2 = {L^3 B \over \sqrt{d}}$.

The entropy density is
\bea\label{cf}
{S\over V} =  {1\over 4G_{d+1}}\left({r_+\over L}\right)^{d-1}
\eea
which at extremality becomes
\bea\label{cg}
{S\over V}={1\over 4G_{d+1}}\left({LB\over \sqrt{d}}\right)^{{d-1\over 2}}~.
\eea
This entropy  is proportional to the number of zero modes of massless
fermions on $T^{d-1}$ in the presence of magnetic flux.

The energy density can be worked out by integrating $ E=\int TdS$ at fixed $B$, and gives
\bea\label{ch}
{E\over V} = {(d-1)\over 16\pi G_{d+1}} {M\over L^{d-1}}~.
\eea
At extremality this becomes
\bea\label{ci}
{E\over V} = {1\over 4\pi G_{d+1}}\left({d-1 \over 4-d}\right) \left({LB\over \sqrt{d}}\right)^{{d\over 2}}{1\over L}~.
\eea
Surprisingly, this is negative for $d>4$.  This does not imply an instability in the
theory, since it is not meaningful to compare this energy against that for $B=0$, as
the geometries have different asymptotics.

\subsection{$d$ even}

We let the field strength fill the directions $x^{1}, x^{2}, \cdots , x^{d-2}$ and denote the ``left over" spatial direction  by $y$.   By rotating and scaling coordinates, and restricting to the symmetric case of equal skew-eigenvalues,  we can
take $F_{12}=F_{34}=\ldots =B$ and  $F_{iy}=0$.   The metric ansatz is
\bea\label{cj}
ds^2 = -U(r)dt^2 +{dr^2\over U(r)} + e^{2V(r)}\Big((dx^1)^2 + \cdots + (dx^{d-2})^2\Big) +e^{2W(r)}dy^2~.
\eea

The Einstein equations can be reduced to
\bea\label{ck}
&& U(V''-W'')+\Bigg(U'+U\Big((d-2)V'+W'\Big)\Bigg)(V'-W')=-2B^2 e^{-4V} \\ \no
&&(d-2)(V''+V'^2)+W''+W'^2 =0  \\ \no
&& (d-2) \bigg (U'V' +2UV'W' +(d-3)UV'^2 \bigg ) +U'W' ={d(d-1) \over L^2} -(d-2)B^2 e^{-4V}
\eea
These equations admit a BTZ$ \times T^{d-2}$   solution
\bea\label{cl}
ds^2 =&-& (d-1)\left({r^2-r_+^2\over L^2}\right)dt^2 + {1 \over (d-1)}\left({L^2 \over r^2-r_+^2}\right)dr^2  +(d-1){r^2 \over L^2}dy^2 \\ \no
 &+&{LB \over \sqrt{d-1}}\Big( (dx^1)^2 + \cdots (dx^{d-2})^2\Big)~.
\eea
The AdS$_3$ radius is $l = {L \over \sqrt{d-1}}$.   The $D=3$ Newton constant is
\bea\label{cm}
{1\over G_3} = \left({LB \over \sqrt{d-1}}\right)^{d-2\over 2}V_{d-2} {1\over G_{d+1}}
\eea
where $V_{d-2}$ denotes the coordinate volume.
The Brown-Henneaux central charge is
\bea\label{cn}
c = {3l \over 2G_3} = {3\over 2} {1 \over \sqrt{d-1}}\left({B \over \sqrt{d-1}L}\right)^{d-2\over 2}V_{d-2} {L^{d-1}\over G_{d+1}}
\eea
This central charge is proportional to that which would arise from massless
fermions on $T^{d-2}$ in the presence of magnetic flux.
The low temperature entropy is
\bea\label{co}
S\approx {\pi \over 3} c T L_y = {\pi\over 2} \left({B \over \sqrt{d-1}L}\right)^{d-2\over 2}{L^{d-1} \over G_{d+1}} {T  \over \sqrt{d-1}}V_{d-1}~.
\eea

The interpolating solutions can be found by numerical integration of (\ref{ck}).

\section{Discussion}
\setcounter{equation}{0}

In this work we have constructed magnetic brane solutions in AdS,  using a combination of analytical and numerical methods.  We mainly focussed on the AdS$_5$ case, and noted
that the black brane thermodynamics agrees surprisingly well  with that of free
$\cN=4$ SYM theory.  Although we framed our discussion in terms of $\cN=4$ SYM theory,
it is worth noting that our solutions apply equally well to the much larger class of
$\cN=1$ superconformal field theories that have a gravity dual described by Einstein-Maxwell theory with a negative cosmological constant.   As shown in
\cite{Buchel:2006gb,Gauntlett:2006ai,Gauntlett:2007ma} this class includes all such theories with
a dual IIB or M-theory description, as all these theories admit a consistent truncation to
$D=5$ minimal gauged supergravity.

A natural extension of this work  is to add nonzero charge and momentum density to these solutions \cite{InProgress}.  This  leads to AdS$_3$ being replaced by a charged, rotating,
BTZ black hole.  The addition of a charge density leads to a nonzero current flow; this can
be seen in $\cN=4$ SYM theory from the fact that there is a left-right asymmetry in the
effective $D=1+1$ CFT.  This is in turn related to the triangle anomaly, and presumably
gives a microscopic explanation for some of the effects noted in \cite{Son:2009tf}.

It would be interesting to study the transport properties of these solutions.  For
that purpose, it  clearly would  be desirable to have an analytic solution available, and in
the appendix we report on our efforts in that direction.

\bigskip\bigskip

\noindent { {\Large \bf Acknowledgments}}

\bigskip

\noindent  This work was
supported in part by NSF grant PHY-07-57702.    We thank the KITP for hospitality during the
``Quantum Criticality and the AdS/CFT Correspondence" program.

\appendix

\section{Towards Exact Solutions}
\setcounter{equation}{0}

It would clearly be valuable to obtain exact analytic solutions to the reduced
equations (\ref{bda}) for the general magnetic brane, and for their boost-invariant
zero temperature limit in (\ref{bg}). The knowledge of an analytic solution
greatly facilitates the study, for example, of the spectrum and dynamics
of small fluctuations. Thus far, we have not succeeded in completely solving
either one of these equations in all generality.

In this Appendix, we shall present two partial analytic solutions of (\ref{bda}) and
(\ref{bg}). The first consists of the regime of large magnetic field, specifically
when the $B^2 e^{-4V}$ terms dominate the constant terms on the rhs of
(\ref{bda}); this problem will be solved completely analytically. The second consists
of the zero temperature regime, where boost invariance in the $x^3$ direction can be assumed; this problem will be reduced, by quadratures, to the solutions of a single
first order ODE. So far, we have not succeeded in solving this remaining ODE.
 In this appendix, we shall provide the derivations of these analytic results.

\subsection{The  large $B$ case}

When $B^2 e^{-4V}$ is large, we ignore the constant terms on the rhs of
(\ref{bda}). The fourth equation is the constraint, which is $r$-independent
in view of the first three equations, and may be enforced as an initial datum.
Using the constraint equation, we may eliminate the $B^2 e^{-4V}$ term
from the first three equations. The resulting equations involve only
the functions $u,v,w$, defined by (the sign and numerical factors are for
later convenience),
\bea
\label{uvw}
u  \equiv  -U'/(6U) \hskip 0.8in v \equiv -V'/3 \hskip 0.8in  w \equiv -W'/3
\eea
and not the actual functions $U,V,W$, and become,
\bea
\label{xvw1}
u' & = & 6 u^2 + 10 uv + 5 uw + 2 v^2 + 4 vw
\no \\
v' & = & -2 uv - 4 uw + 2 v^2 - 5 vw
\no \\
w' & = & 4 uv + 8 uw + 2 v^2 + 3 w^2 + 10 vw
\eea
In the two independent ratios $u'/v'$ and $w'/v'$, the derivative
with respect to $r$ is effectively traded in for a derivative with respect to $v$.
The right hand sides are homogeneous in $u,v,w$ of degree 0, and
may be expressed solely in terms of the ratios,
\bea
\label{albet}
\a \equiv u/v \hskip 1in \b \equiv w/ v
\eea
and we obtain,
\bea
\label{albet1}
v { d\a \over dv} & = &
{4 \a^2 \b + 8  \a^2 + 10 \a \b + 8 \a + 4 \b +2 \over - 4 \a \b - 2 \a - 5 \b +2}
\no \\
v { d \b \over dv} & = &
{ 4 \a \b^2 + 8 \b^2 + 10 \a \b + 4 \a + 8 \b + 2 \over - 4 \a \b - 2 \a - 5 \b +2}
\eea
Taking  the ratio of these equations in turn gives an ordinary
first order differential equation,
\bea
\label{albet2}
{ d \b \over d \a} = { 4 \a \b^2 + 8 \b^2 + 10 \a \b + 4 \a + 8 \b + 2
\over 4 \a^2 \b + 8  \a^2 + 10 \a \b + 8 \a + 4 \b +2}
\eea
Equation (\ref{albet2}) may be solved by considering the
ratio $d(\a - \b) /d(\a + \b) = (\a-\b)/\lambda$ in terms of the independent
variable $\lambda \equiv \a + \b +1$. The general solution is given by,
\bea
\a - \b = c \lambda
\eea
where $c$ is an arbitrary real integration constant. Using this solution back into
(\ref{albet1}), we derive $v$ by quadrature, and by using (\ref{albet}), we find
$u$ and $w$ as well,
\bea
\label{xvw3}
u(\lambda) & = & \half v_0 \Big \{ (1+c)\lambda -1 \Big \} \lambda ^{- 3/2}
(\lambda - \lambda _+)^{\g_+} (\lambda - \lambda _-)^{\g_-}
\no \\
v(\lambda) & = & v_0  \lambda ^{- 3/2}
(\lambda - \lambda _+)^{\g_+} (\lambda - \lambda _-)^{\g_-}
\no \\
w(\lambda) & = & \half v_0 \Big \{ (1-c)\lambda -1 \Big \} \lambda ^{- 3/2}
(\lambda - \lambda _+)^{\g_+} (\lambda - \lambda _-)^{\g_-}
\eea
where $ v_0$ is a real integration constant, and the constants $\lambda _\pm$
and  $\g_\pm$ are given by,
\bea
\lambda _\pm & \equiv &  { - 3 \pm  \sqrt{ 12 - 3 c^2} \over 1-c^2}
\no \\
\g _\pm & = & \pm
{ 3(c+3)  + 3 \lambda _\pm ^{-1} \over 4 \sqrt{ 12 - 3 c^2}}
\eea
Finally, we use (\ref{xvw1}) to obtain also $r$ in terms of $\lambda$, and we find,
\bea
v_0 (1-c^2) r(\lambda)  = \int _{\lambda _0} ^\lambda d \lambda' \, (\lambda') ^\half
(\lambda' - \lambda _+)^{-1-\g_+ } (\lambda' - \lambda _-)^{-1- \g_-}
\eea
where $\lambda _0$ is a real integration constant. For the special case of
the boost invariant solution, for which $c=0$,  the data work out as follows,
$\lambda _\pm = - 3 \pm 2 \sqrt{3}$, and $4 \g _\pm = \pm 2  \sqrt{3} + 1$.

\subsection{The boost-invariant case}

Boost invariance in the $x^3$-direction requires $U=3 e^{2 W}$, and reduces
the equations (\ref{bda}) to,
\bea
\label{red6}
2 V'' + 2(V')^2 + W'' +(W')^2 & = & 0
\no \\
V''-W'' + 2 (V')^2 - 3 (W')^2 + V'W' & = &- {2 \over 3} B^2 e^{-4V-2W}
\no \\
(V')^2 + (W')^2 + 4 V'W' & = & { 2 \over L^2} e^{-2W} -{1 \over 3} B^2 e^{-4V-2W}
\eea
It is readily checked that the $r$-derivative of the constraint vanishes in
view of the first to equations of (\ref{red6}). Thus, the constraint may again be
imposed as initial conditions. To solve the system, we concentrate of
the first two equations of (\ref{red6}).

\sm

Taking the derivative of the second equation, and eliminating the $B^2 e^{-4 V-2W}$
between the original equation and its derivative gives an equation that only
involves $V'$ and $W'$ and its derivatives, but not the original fields $V,W$.
We introduce again the notations of (\ref{uvw}) for $v$ and $w$, eliminate
$w''$ using the derivative of the first equation in (\ref{red6}), so that
the final two independent equations become,
\bea
\label{vw2}
2v' + w' & = & 6 v^2 + 3 w^2
\no \\
- v'' + 8 vv' - 4 ww' + v'w+ vw' & = &
3 (4 v + 2 w) ( - v' + 4 v^2 - 2 w^2 + vw)
\eea
To render the system first order in derivatives on $v$,
we introduce an auxiliary variable $y$, and postulate that the first
derivatives of $v$ and $y$ be quadratic functions
of $v,w,y$. Clearly, $y$ is not unique, but is defined up to
linear transformations on $y$ of the form, $y \to s y + a v+bw$, for $s,a,b$
arbitrary real parameters. One readily establishes that such a system is given by,
\bea
\label{wvy}
v' & = & 3 v^2 - y(v-w)
\no \\
w ' & = & 3w^2 +2  y(v-w)
\no \\
y' & = & 6 v^2 + 3 y^2 + 21 vw + 9 vy + 12 wy
\eea
One verifies that, if $w,v,y$ satisfy (\ref{wvy}), then $v,w$ satisfy  (\ref{vw2}).
The $AdS_5$ solution has $v=w=1/(3r)$ and $ y \sim 1/r$, while the
$AdS_3$ solution has $v=y=0$ and $w= 1/(3r)$.

\sm

The system may be reduced to a single first order ordinary ODE by taking
pairwise ratios $v'/w'$ and $y'/w'$, and using the ratios $\a \equiv v/w$ and
$\b \equiv y/w$. The resulting equations are,
\bea
w { d \a \over dw} & = & {(\a-1) (3 \a - \b - 2 \a \b) \over 3 + 2 \b (\a - 1)}
\no \\
w { d \b \over dw} & = & {  6\a^2 + 5 \b^2 + 21 \a + 9 \b + 9 \a \b
- 2 \a \b^2  \over 3 + 2 \b (\a - 1)}
\eea
Taking the ratio, we get a single first order ODE between $\a$ and $\b$,
\bea
\label{ab6}
{ d \a \over d \b} = { (\a-1) ( 3 \a  - \b - 2 \a \b) \over
6\a^2 + 5 \b^2 + 21 \a + 9 \b + 9 \a \b  - 2 \a \b^2}
\eea
The solution $\a=1$ corresponds to $AdS_5$, while $\a = \b=0$
corresponds to $AdS_3 \times T^2$. Equation (\ref{ab6}) is of the Darboux type of order $m=2$ according
to \cite{Ince}, and of the Abel second kind type according to
\cite{PZ}. No general solutions are known for either type of
equations, and (\ref{ab6}) does not fit into any of the known categories
of solvable special cases. Either way, we have not succeeded in
integrating equation (\ref{ab6}) analytically.


\begin{thebibliography}{99}

\bibitem{Hartnoll:2007ai}
  S.~A.~Hartnoll and P.~Kovtun,
  ``Hall conductivity from dyonic black holes,''
  Phys.\ Rev.\  D {\bf 76}, 066001 (2007)
  [arXiv:0704.1160 [hep-th]].

\bibitem{Hartnoll:2007ih}
  S.~A.~Hartnoll, P.~K.~Kovtun, M.~Muller and S.~Sachdev,
  ``Theory of the Nernst effect near quantum phase transitions in condensed
  matter, and in dyonic black holes,''
  Phys.\ Rev.\  B {\bf 76}, 144502 (2007)
  [arXiv:0706.3215 [cond-mat.str-el]].

\bibitem{Hartnoll:2007ip}
  S.~A.~Hartnoll and C.~P.~Herzog,
  ``Ohm's Law at strong coupling: S duality and the cyclotron resonance,''
  Phys.\ Rev.\  D {\bf 76}, 106012 (2007)
  [arXiv:0706.3228 [hep-th]].

\bibitem{Albash:2008eh}
  T.~Albash and C.~V.~Johnson,
  ``A Holographic Superconductor in an External Magnetic Field,''
  JHEP {\bf 0809}, 121 (2008)
  [arXiv:0804.3466 [hep-th]].

\bibitem{Buchbinder:2008dc}
  E.~I.~Buchbinder, A.~Buchel and S.~E.~Vazquez,
  ``Sound Waves in (2+1) Dimensional Holographic Magnetic Fluids,''
  JHEP {\bf 0812}, 090 (2008)
  [arXiv:0810.4094 [hep-th]].

\bibitem{Hansen:2008tq}
  J.~Hansen and P.~Kraus,
  ``Nonlinear Magnetohydrodynamics from Gravity,''
  JHEP {\bf 0904}, 048 (2009)
  [arXiv:0811.3468 [hep-th]].

\bibitem{Caldarelli:2008ze}
  M.~M.~Caldarelli, O.~J.~C.~Dias and D.~Klemm,
  ``Dyonic AdS black holes from magnetohydrodynamics,''
  JHEP {\bf 0903}, 025 (2009)
  [arXiv:0812.0801 [hep-th]].

\bibitem{Hansen:2009xe}
  J.~Hansen and P.~Kraus,
  ``S-duality in AdS/CFT magnetohydrodynamics,''
  arXiv:0907.2739 [hep-th].

\bibitem{Denef:2009yy}
  F.~Denef, S.~A.~Hartnoll and S.~Sachdev,
  ``Quantum oscillations and black hole ringing,''
  arXiv:0908.1788 [hep-th].

\bibitem{Albash:2009wz}
  T.~Albash and C.~V.~Johnson,
  ``Holographic Aspects of Fermi Liquids in a Background Magnetic Field,''
  arXiv:0907.5406 [hep-th].

\bibitem{Basu:2009qz}
  P.~Basu, J.~He, A.~Mukherjee and H.~H.~Shieh,
  ``Holographic Non-Fermi Liquid in a Background Magnetic Field,''
  arXiv:0908.1436 [hep-th].


\bibitem{Son:2009tf}
  D.~T.~Son and P.~Surowka,
  ``Hydrodynamics with Triangle Anomalies,''
  arXiv:0906.5044 [hep-th].

\bibitem{Chamseddine:1999xk}
  A.~H.~Chamseddine and W.~A.~Sabra,
  ``Magnetic strings in five dimensional gauged supergravity theories,''
  Phys.\ Lett.\  B {\bf 477}, 329 (2000)
  [arXiv:hep-th/9911195].

\bibitem{Klemm:2000nj}
  D.~Klemm and W.~A.~Sabra,
  ``Supersymmetry of black strings in D = 5 gauged supergravities,''
  Phys.\ Rev.\  D {\bf 62}, 024003 (2000)
  [arXiv:hep-th/0001131].

\bibitem{Sabra:2002xy}
  W.~A.~Sabra,
  ``Magnetic branes in d-dimensional AdS Einstein-Maxwell gravity,''
  Phys.\ Lett.\  B {\bf 545}, 175 (2002)
  [arXiv:hep-th/0207128].


\bibitem{Cucu:2003yk}
  S.~Cucu, H.~Lu and J.~F.~Vazquez-Poritz,
  ``Interpolating from AdS(D-2) x S(2) to AdS(D),''
  Nucl.\ Phys.\  B {\bf 677}, 181 (2004)
  [arXiv:hep-th/0304022].


\bibitem{Brihaye:2007vm}
  Y.~Brihaye, E.~Radu and C.~Stelea,
  ``Black strings with negative cosmological constant: Inclusion of electric
  charge and rotation,''
  Class.\ Quant.\ Grav.\  {\bf 24}, 4839 (2007)
  [arXiv:hep-th/0703046].


\bibitem{Bernamonti:2007bu}
  A.~Bernamonti, M.~M.~Caldarelli, D.~Klemm, R.~Olea, C.~Sieg and E.~Zorzan,
  ``Black strings in AdS$_5$,''
  JHEP {\bf 0801}, 061 (2008)
  [arXiv:0708.2402 [hep-th]].

\bibitem{Maldacena:2000mw}
  J.~M.~Maldacena and C.~Nunez,
  ``Supergravity description of field theories on curved manifolds and a no  go
  theorem,''
  Int.\ J.\ Mod.\ Phys.\  A {\bf 16}, 822 (2001)
  [arXiv:hep-th/0007018].


\bibitem{Banados:1992wn}
  M.~Banados, C.~Teitelboim and J.~Zanelli,
  ``The Black hole in three-dimensional space-time,''
  Phys.\ Rev.\ Lett.\  {\bf 69}, 1849 (1992)
  [arXiv:hep-th/9204099].

\bibitem{Henningson:1998gx}
  M.~Henningson and K.~Skenderis,
  ``The holographic Weyl anomaly,''
  JHEP {\bf 9807}, 023 (1998)
  [arXiv:hep-th/9806087].



\bibitem{Balasubramanian:1999re}
  V.~Balasubramanian and P.~Kraus,
  ``A stress tensor for anti-de Sitter gravity,''
  Commun.\ Math.\ Phys.\  {\bf 208}, 413 (1999)
  [arXiv:hep-th/9902121].


\bibitem{Gauntlett:2003fk}
  J.~P.~Gauntlett and J.~B.~Gutowski,
  ``All supersymmetric solutions of minimal gauged supergravity in five
  dimensions,''
  Phys.\ Rev.\  D {\bf 68}, 105009 (2003)
  [Erratum-ibid.\  D {\bf 70}, 089901 (2004)]
  [arXiv:hep-th/0304064].

\bibitem{Brown:1986nw}
  J.~D.~Brown and M.~Henneaux,
  ``Central Charges in the Canonical Realization of Asymptotic Symmetries: An
  Example from Three-Dimensional Gravity,''
  Commun.\ Math.\ Phys.\  {\bf 104}, 207 (1986).

\bibitem{Ortaggio:2007hs}
  M.~Ortaggio, J.~Podolsky and M.~Zofka,
  ``Robinson-Trautman spacetimes with an electromagnetic field in higher
  dimensions,''
  Class.\ Quant.\ Grav.\  {\bf 25}, 025006 (2008)
  [arXiv:0708.4299 [gr-qc]].



\bibitem{Buchel:2006gb}
  A.~Buchel and J.~T.~Liu,
  ``Gauged supergravity from type IIB string theory on Y(p,q) manifolds,''
  Nucl.\ Phys.\  B {\bf 771} (2007) 93
  [arXiv:hep-th/0608002].

\bibitem{Gauntlett:2006ai}
  J.~P.~Gauntlett, E.~O Colgain and O.~Varela,
  ``Properties of some conformal field theories with M-theory duals,''
  JHEP {\bf 0702} (2007) 049
  [arXiv:hep-th/0611219].

\bibitem{Gauntlett:2007ma}
  J.~P.~Gauntlett and O.~Varela,
  ``Consistent Kaluza-Klein Reductions for General Supersymmetric AdS
  Solut<ions,''
  Phys.\ Rev.\  D {\bf 76} (2007) 126007
  [arXiv:0707.2315 [hep-th]].

\bibitem{InProgress}
E.~D'Hoker and P.~Kraus, work in progress.

\bibitem{Ince}
E.L. Ince, {\sl Ordinary Differential Equations},
Dover 1956, page 29.

\bibitem{PZ}
A.D. Polianin, V.F. Zaitsev, {\sl Handbook of exact solutions
for ordinary differential equations}, Chapman and Hall/CRC, page 10.


\end{thebibliography}
\end{document}